\documentclass[10pt,twoside,twocolumn,english,aps,manuscript,aps,preprint,show pacs,prb]{revtex4}
\usepackage[T1]{fontenc}
\usepackage[latin9]{inputenc}
\usepackage{babel}
\usepackage{array}
\usepackage{longtable}
\usepackage{textcomp}
\usepackage{amsmath}
\usepackage{graphicx}
\usepackage{esint}
\usepackage[unicode=true,pdfusetitle,
 bookmarks=true,bookmarksnumbered=false,bookmarksopen=false,
 breaklinks=false,pdfborder={0 0 1},backref=false,colorlinks=false]
 {hyperref}
\usepackage{breakurl}

\makeatletter

\DeclareRobustCommand{\greektext}{%
  \fontencoding{LGR}\selectfont\def\encodingdefault{LGR}}
\DeclareRobustCommand{\textgreek}[1]{\leavevmode{\greektext #1}}
\DeclareFontEncoding{LGR}{}{}
\DeclareTextSymbol{\~}{LGR}{126}
\newcommand{\lyxmathsym}[1]{\ifmmode\begingroup\def\b@ld{bold}
  \text{\ifx\math@version\b@ld\bfseries\fi#1}\endgroup\else#1\fi}

\providecommand{\tabularnewline}{\\}

\@ifundefined{textcolor}{}
{%
 \definecolor{BLACK}{gray}{0}
 \definecolor{WHITE}{gray}{1}
 \definecolor{RED}{rgb}{1,0,0}
 \definecolor{GREEN}{rgb}{0,1,0}
 \definecolor{BLUE}{rgb}{0,0,1}
 \definecolor{CYAN}{cmyk}{1,0,0,0}
 \definecolor{MAGENTA}{cmyk}{0,1,0,0}
 \definecolor{YELLOW}{cmyk}{0,0,1,0}
}

\makeatother

\begin{document}

\title{Cluster spin glass behavior in geometrically frustrated Zn$_{\text{3}}$V$_{\text{3}}$O$_{\text{8}}$ }

\author{T. Chakrabarty}

\email{tanmoyc@iitb.ac.in}

\affiliation{Department of Physics, IIT Bombay, Powai, Mumbai 400076, India }

\author{A. V. Mahajan}

\email{mahajan@phy.iitb.ac.in}

\affiliation{Department of Physics, IIT Bombay, Powai, Mumbai 400076, India}

\author{S. Kundu}

\affiliation{Department of Physics, IIT Bombay, Powai, Mumbai 400076, India}
\begin{abstract}
We report the bulk magnetic properties of a yet unexplored vanadium-based
multi-valenced spinel system, Zn$_{3}$V$_{3}$O$_{8}$. A Curie-Weiss
fit of our \textit{dc} magnetic susceptibility $\chi(T)$ data in
the temperature region of $140\lyxmathsym{\textminus}300$ K yields
a Curie constant $C=0.75$ cm$^{\text{3}}$K/mole V, $\theta_{CW}=-370$
K. We have observed a splitting between the zero field cooled ZFC
and field cooled FC susceptibility curves below a temperature $T_{irr}$
of about $6.3$ K. The value of the \textquoteleft{}frustration parameter\textquoteright{}
($\frac{\left|\theta_{\mathrm{CW}}\right|}{T_{\mathrm{N}}}\sim100$)
suggests that the system is strongly frustrated. From the \textit{ac}
susceptibility measurements we find a logarithmic variation of freezing
temperature ($T_{f}$) with frequency $\nu$ attesting to the formation
of a spin glass below $T_{f}$. However, the value of the characteristic
frequency obtained from the Vogel-Fulcher fit suggests that the ground
state is closer to a cluster glass rather than a conventional spin
glass. We explored further consequences of the spin glass behavior
and observed aging phenomena and memory effect (both in ZFC and FC).
We found that a positive temperature cycle erases the memory, as predicted
by the hierarchical model. From the heat capacity $C_{P}$ data, a
hump-like anomaly was observed in $C_{P}/T$ at about $3.75$ K. Below
this temperature the magnetic heat capacity shows a nearly linear
dependence with $T$ which is consistent with the formation of a spin
glass state below $T_{f}$ in Zn$_{3}$V$_{3}$O$_{8}$.
\end{abstract}


\maketitle

\section{introduction}

A great deal of research has been done on the ground state properties
of geometrically frustrated antiferromagnets during the last decade.\cite{J.E.Greedan-Geometrically frustrated magnetic materials-2000,Physics Today-2006 A.P.Ramirez geometric frustration}
Cubic spinels (AB$_{2}$O$_{4}$) and pyrochlores (A$_{2}$B$_{2}$O$_{7}$)
with magnetic ions at the B-sites are important in this respect due
to the formation of a geometrically frustrated corner-shared tetrahedral
network among themselves. Na$_{4}$Ir$_{3}$O$_{8}$ is a famous compound
of the spinel family found in recent times where Ir and Na at the
B-sites have distinct positions and the Ir ions form corner shared
triangles in three dimensions. This has been dubbed as the hyperkagome
lattice and is proposed to have a quantum spin liquid ground state.
In some cases where there is deviation from the frustrated geometry,
the resulting non-uniform interaction can lead to the formation of
a spin glass SG state which is an exotic example of a nonergodic state
marked by several other physical properties such as slow dynamics,
nonexponential decay, aging effect, memory effect, etc. \cite{Binder Young SG open questions,memory and chaos effect in SG }.
In pyrochlores and spinels, spin glass behavior is often observed
where the requirement of non uniform interaction can be readily satisfied.
The nonequilibrium spin glass state arises from frustration due to
competing magnetic interactions among the spins as well as disorder
capable of pinning the spins. In the last few years, spin-glass-like
nonequilibrium dynamics and time-dependent behavior have been observed
in several magnetic systems, where the basic building blocks responsible
for this \textquotedblleft{}glassy\textquotedblright{} behavior are
spin clusters or a bigger spin entities rather than atomic spins.\cite{memory effect in GdCu,JPCM Nd5Ge3}
Till now this sort of behavior has been prominently observed in manganites,
\cite{PRB SG Pr0.7Ca0.3MnO3,SG manganites perovskites} cobaltites,
\cite{Spin-glass behavior in Ni-doped La1.85Sr0.15CuO4,cluster SG cobaltites,aging effect in RSG layered manganites}
intermetallic alloys, \cite{JPCM Nd5Ge3,cluster SG in Gd5Ge4,Ce1-xErx Fe2 EPL}
other oxides, \cite{Ca3CoRhO6,LaCo0.5Ni0.5O3 PSG PRB Vishwanathan}
as also in magnetic nanoparticles\cite{memory effect in nanoparticle}
where the frozen state is not the conventional spin glass but is rather
referred to as a cluster glass since the basic building block is bigger
than the simple atomic spins. Till date very few vanadium-based oxides
have been placed in this category.

We were in the quest for new spinel compounds with magnetic ions at
the B-sites, with the intention of unraveling novel magnetic properties
arising due to geometric frustration. In this paper, we report our
studies on the yet unexplored system Zn$_{\text{3}}$V$_{\text{3}}$O$_{\text{8}}$
via magnetization and heat capacity. Here, although there are magnetic
V atoms at the B sites, the network is diluted due to the presence
of nonmagnetic Zn, a third of which reside at the B site. Our magnetization
data are consistent with the presence of two V$^{3+}$($S=1$) and
one V$^{4+}$($S=1/2$) ions per formula unit. A large, negative,
Curie-Weiss temperature $\theta_{CW}$ = $-370$ K is found which
is indicative of strong antiferromagnetic (AF) interactions. A difference
between the zero field cooled (ZFC) and field cooled (FC) susceptibilities
is observed below about $6.3$ K. The heat capacity also has a low
temperature anomaly. Further, \textit{ac} susceptibility measurements
corroborate the formation of a spin glass state, more specifically,
a cluster spin glass. To study the dynamics of this glassy state below
the freezing temperature ($T_{f}$), we carried out further experiments
and observed magnetic relaxation, memory and aging phenomena which
are thought to be typical characteristics of spin glass dynamics.
During the study of memory effects, we observed that a positive heat
cycle erases the previous memory and initializes the relaxation again
as prescribed by the hierarchical model.\cite{memory effect in nanoparticle,Hierarchical model EPL}

\section{experimental details and structural information}

\begin{figure}
\begin{centering}
\includegraphics[scale=0.34]{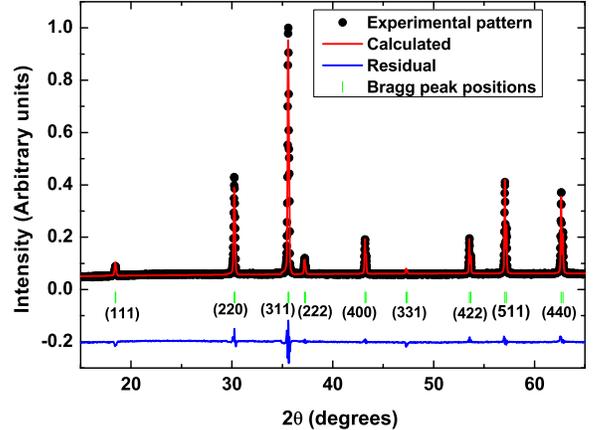}
\par\end{centering}

\caption{\label{fig:1xrd}Powder diffraction pattern of Zn$_{\text{3}}$V$_{\text{3}}$O$_{\text{8}}$
is shown along with its Bragg peak positions; The black points are
the experimental data, the red solid line is the calculated refinement
pattern, the blue solid line is the residual data and the green markers
are the Bragg peak positions for\textit{ F d -3 m-s} space group. }
\end{figure}

Zn$_{\text{3}}$V$_{\text{3}}$O$_{\text{8}}$ was prepared by standard
solid state reaction methods. First we prepared V$_{2}$O$_{3}$ by
reducing V$_{2}$O$_{5}$ (Aldrich\textemdash{}99.99\%) in hydrogen
atmosphere at $650^{\circ}$C for $16$ hours and VO$_{2}$ by mixing
V$_{2}$O$_{3}$ with V$_{2}$O$_{5}$ in a $1:1$ molar ratio, pelletizing
and firing in dynamical vacuum (better than $10^{-5}$ mbar) at $800^{\circ}$C
for 24 hours. In the first step we pelletized a mixture of ZnO (Aldrich
99.99\% purity, dried at $140^{\circ}$C to remove moisture), V$_{2}$O$_{3}$,
and VO$_{2}$ which was then sealed in an evacuated quartz tube. This
was then fired at $650^{\circ}$C for $24$ hours. After regrinding
and repelletizing, the sample was again fired in an evacuated quartz
tube at $850^{\circ}$C. X-ray diffraction (xrd) patterns were collected
with a PANalytical x-ray diffractometer using Cu K$_{\alpha}$radiation
($\lambda=\mbox{1.54182 }\textrm{\ensuremath{\mathrm{\AA}}}$). From
the Rietveld refinement of the xrd pattern (see Fig. \ref{fig:1xrd})
the lattice parameter was found to be $8.402$ Å in the \textit{F
d -3 m} space group. The results of the refinements are shown in Table
\ref{atomic positions} below. The goodness of the Rietveld refinement
is defined by the following parameters. R$_{p}$= $3.94$\%, R$_{wp}$
= $5.42$ \%, R$_{exp}$= $2.33$\%, and $\lyxmathsym{\textgreek{q}}$$^{2}$
= $5.39$. 

\begin{table}
\caption{\label{atomic positions}Atomic positions in Zn$_{3}$V$_{3}$O$_{8}$}
\begin{longtable}{|c|c|c|c|c|}
\hline 
Atoms & \multicolumn{1}{c}{} & \multicolumn{1}{c}{Co-ordinates} &  & Occupancy\tabularnewline
\cline{2-5} 
 & x/a & y/b & z/c & \tabularnewline
\hline 
Zn1(8a) & $0.000$ & $0.000$ & $0.000$ & $0.985$\tabularnewline
\hline 
Zn2(16d) & $0.625$ & $0.625$ & $0.625$ & $0.225$\tabularnewline
\hline 
V(16d) & $0.625$ & $0.625$ & $0.625$ & $0.775$\tabularnewline
\hline 
O1(32e) & $0.734$ & $0.734$ & $0.734$ & $1.000$\tabularnewline
\hline 
\end{longtable}
\end{table}

\begin{figure}
\includegraphics[scale=0.48]{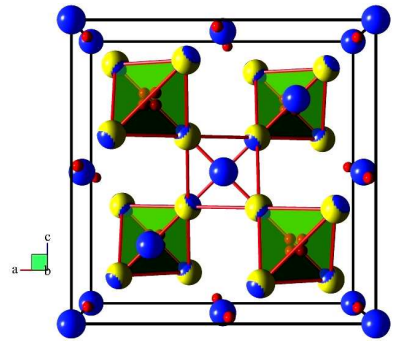}

\caption{\label{fig:2unit cell}The unit cell of Zn$_{\text{3}}$V$_{\text{3}}$O$_{\text{8}}$.
The Zn$^{2+}$, V$^{4+}$/V$^{3+}$, O$^{2-}$ions are shown in blue,
yellow, and red color, respectively. The B-sites form a corner-shared
tetrahedral network consisting of V$^{3+}$, V$^{4+},$ and Zn$^{2+}$ions. }
\end{figure}

In the structure of Zn$_{\text{3}}$V$_{\text{3}}$O$_{\text{8}}$
the B-sites are shared by zinc (Zn$^{2+}$ has an ionic radius of
$0.74$ Å) and vanadium (both V$^{4+}$ and V$^{3+}$) in a $1:3$
ratio to form a corner-shared tetrahedral network {[}see Fig. \ref{fig:2unit cell}{]}.
The V$^{4+}$ ion ($S=1/2$ and ionic radius $0.58$ Å) differs slightly
in size from the V$^{3+}$ ion ($S=1$ and ionic radius $0.64$ Å).
Due to this difference in ionic size the corner-shared tetrahedral
network may be distorted and due to the difference in the value of
spin the various 3\textit{d}-3\textit{d} nearest neighbour magnetic
interactions will likely be different. In addition to this, due to
the random/statistical occupation of the B sites by Zn and V, there
are going to be missing magnetic ions in the triangular network. Therefore,
Zn$_{\text{3}}$V$_{\text{3}}$O$_{\text{8}}$ can be regarded as
a bond-disordered geometrically frustrated antiferromagnet, as for
LiCrMnO$_{4}$.\cite{LiCrMnO4 sg PRB} This disruption/dilution of
the corner-shared tetrahedral network serves as an additional source
of disorder and is likely to lead to relieving of frustration and
an eventual spin glass/frozen state. 

The temperature dependence of magnetization $M$, its relaxation,
memory effect, aging phenomenon and several other dynamic properties
were measured in the temperature range 2-300 K using a SQUID VSM from
Quantum Design. The temperature dependence of heat capacity has also
been measured in the temperature range of 2-250 K using the heat capacity
attachment of a Quantum Design PPMS.

\section{results and discussions}

\subsection{Magnetic measurementsthe}

\subsubsection{\textbf{dc magnetization and ac susceptibility}}

\begin{figure}
\includegraphics[scale=0.32]{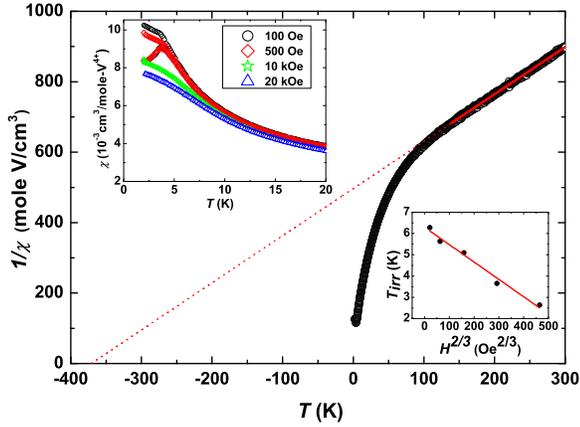}

\caption{\label{fig:3ZFC-FC}Temperature dependence of inverse susceptibility
$1/(\chi-\chi_{0})$ is shown for $H=100$ Oe. The red line shows
the Curie-Weiss fit in the temperature range of $140-300$ K and the
dotted line is its extrapolation.The left inset shows the bifurcation
between the ZFC and FC curves for various fields. The right inset
shows the de Almeida\textendash{}Thouless line, plotted as $T_{irr}$
vs $H^{2/3}$. }
\end{figure}

The reciprocal of the susceptibility is plotted with temperature in
Fig. \ref{fig:3ZFC-FC}. From the Curie-Weiss fit $\chi(T)$ = $\lyxmathsym{\textgreek{q}}$$_{\text{0}}$+
$C/(T-\theta_{\text{CW}})$ in the range $140-300$ K, we get the
$T$-independent susceptibility $\mbox{\textgreek{q}}_{\text{0}}$
= $4.66\times10^{\text{-4}}$ cm$^{\text{3}}$/mole V, the Curie constant
$C=0.75$ cm$^{\text{3}}$K/mole V, and the Curie-Weiss temperature
$\theta_{\text{CW}}=-370$ K. From $\mbox{\textgreek{q}}_{\text{0}}$
= $4.66\times10^{\text{-4}}$ cm$^{\text{3}}$/mole-V we obtain the
Van Vleck susceptibility $\mbox{\ensuremath{\chi}}_{\text{VV}}=\mbox{\textgreek{q}}_{\text{0}}-\chi_{\mathrm{core}}=6.19\times10^{\text{-4}}$
cm$^{\text{3}}$/mole V where $\chi_{\text{core}}$ is the core diamagnetic
susceptibility equal to $-1.53\times10^{-4}$ cm$^{\text{3}}$/ mole
formula unit.\cite{Core susceptibility} The value of the Curie constant
for a $S=1/2$ system (with $g=2$) is expected to be $0.37$5 cm$^{\text{3}}$K/mole
and $1$ cm$^{\text{3}}$K/mole for $S=1$. The Curie constant obtained
from the fit (i.e., $C=0.792$ cm$^{\text{3}}$K/mole V) is consistent
with having two $S=1$ and one $S=1/2$ magnetic ion per formula unit.
In low fields, a difference betwen the ZFC and FC susceptibilities,
suggestive of a spin glass transition, is seen below a $T_{irr}$
of about $6.3$ K. The ZFC and FC curves coalesce in fields of $20$
kOe. A large frustration parameter $\frac{\left|\theta_{\mathrm{CW}}\right|}{T_{\mathrm{N}}}\sim100$
suggests that strong frustration is present in the sytem.\cite{Shahab f>5}
It is found that $T$$_{irr}$ follows the $H^{2/3}$ law, which has
been observed in many spin glass systems such as Nd$_{5}$Ge$_{3}$(Ref.\cite{JPCM Nd5Ge3}),
Nd$_{2}$AgIn$_{3}$(Ref.\cite{Nd5AgIn3 almeida Thouless APL}), and
U$_{2}$IrSi$_{3}$(Ref.\cite{U2IrSi3 high f0 conventional SG}). 

\begin{figure}
\includegraphics[scale=0.35]{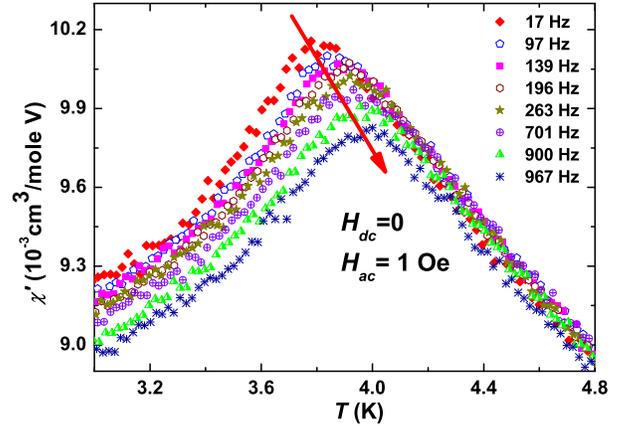}\caption{\label{fig:4ac susceptibility chi'}Temperature dependence of the
real part of $ac$ susceptibility ($\chi'$), measured at different
frequencies with zero external \textit{dc} magnetic field and an $ac$
field of $1$ Oe. The red arrow points to the shift in the spin-glass
temperature ($T_{f}$) with increase in frequency.}
\end{figure}

\begin{figure}

\includegraphics[scale=0.35]{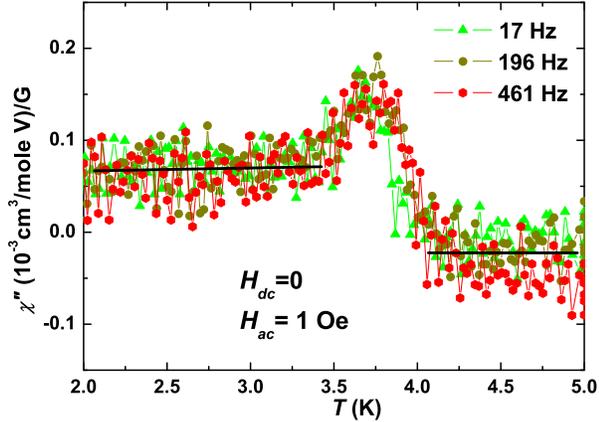}\caption{\label{fig:5 ac susceptibility X"}Temperature dependence of the out-of-phase
part of $ac$ susceptibility ($\chi''$), measured at different frequencies
with zero external \textit{dc} magnetic field and an $ac$ field of
amplitude $1$ Oe.}
\end{figure}

\begin{figure}

\includegraphics[scale=0.32]{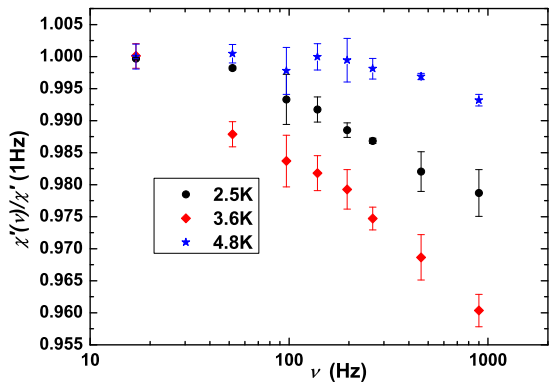}\caption{\label{fig:6log dependence}Normalized real part of $\chi'(T)$ for
Zn$_{\text{3}}$V$_{\text{3}}$O$_{\text{8}}$ as a function of frequency
for temperatures above and below $T_{f}$. The measurements were carried
out in an \textit{ac} field of amplitude $1$ Oe and at zero \textit{dc}
field.}
\end{figure}

Fig. \ref{fig:4ac susceptibility chi'} shows the temperature variations
of the in-phase component of the \textit{ac} magnetic susceptibility
($\chi'$($T$,$\nu$)) measured between $3$ and $5$ K in the frequency
range $17$ \ensuremath{\le} $\omega/2\pi$ \ensuremath{\le} $967$
Hz . The $\chi'$ ($T$,$\nu$) curve displays a peak at $T_{f}\sim3.75$
K and it shifts towards higher temperatures as the frequency $\nu$
changes from $17$ Hz to $967$ Hz. The out of phase component of
\textit{ac} susceptibility $\chi''$($T$,$\nu$) also shows a peak
at $3.75$ K which however hardly shifts with a change in frequency
(see Fig. \ref{fig:5 ac susceptibility X"}). At higher temperatures,
above the freezing temperature $T_{f}$, $\chi''(T)$ is nearly equal
to zero while below $T_{f}$ it has a non zero value. Such behavior
is characteristic of the SG transition and allows us to distinguish
the SG compounds from the disordered AF systems, in which $\chi''(T)$
is constant and remains zero even below the transition temperature.
\cite{Chi" is  not zero below Tf,AuMn canonical Sg Mydosh PRB,URh2Ge2 Chi" not zero below Tf}
In Fig. \ref{fig:6log dependence}, we depict $\chi''(T)$, normalized
to the value at 17 Hz, as a function of frequency for Zn$_{\text{3}}$V$_{\text{3}}$O$_{\text{8}}$,
as an example. At $4.8$ K, i.e., above $T_{f}$, the variation of
$\nu$ over two decades does not influence $\chi'$ in a noticeable
way. Below $T_{f}$, in the SG state, the $\chi'$ exhibits a logarithmic
frequency dependence. This kind of frequency dependence has been predicted
theoretically for a short-range Ising SG\cite{ising sg}, and it has
been observed for several other SG systems\cite{Spin-glass behavior in Ni-doped La1.85Sr0.15CuO4,log dependence  in f Bi0.67Ca0.33MnO3}.
Note that at $3.6$ K, closer to $T_{f}$, the system is more sensitive
to the variation of frequency than at $2.5$ K, which is \textquotedblleft{}deeper\textquotedblright{}
into the frozen state.

The \textit{ac} measurements at various frequencies reveal that the
position of the maximum in the $\chi'(T)$ curve ($T_{f}$), moves
toward higher temperatures and the magnitude of $\chi'$ decreases
with increasing frequency. Such behavior is expected for an SG system.
Now, we will try to determine the category of SG to which Zn$_{\text{3}}$V$_{\text{3}}$O$_{\text{8}}$
belongs. As a raw measure of this frequency dependence, we have calculated
the relative shift of the spin freezing temperature ($\delta T_{f}$
= $\Delta T_{f}/T_{f}$$\Delta log\nu$)\cite{AuMn canonical Sg Mydosh PRB}
and it comes out to be $0.028$. This value of $\delta T_{f}$ indicates
that the sensitivity to the frequency is larger than that for canonical
spin glasses such as CuMn ($\delta T_{f}=0.005$) (Ref.\cite{Mydosh relative shift in Tf})
and AuMn ($\delta T_{f}$=$0.0045$)(Ref.\cite{AuMn canonical Sg Mydosh PRB}).
It is in fact intermediate between the value of canonical spin glass
system and superparamagnets ( for noninteracting ideal superparamagnetic
systems such as holmium borate ($\alpha$-{[}Ho$_{2}$O$_{3}$(B$_{2}$O$_{3}$){]}),
$\delta T_{f}\sim0.28$)\cite{Mydosh relative shift in Tf} where
the spin glass states appear due to the interaction between the randomly
distributed magnetic clusters; however it is close to that typical
for cluster glass systems. . The sensitivity to frequency strongly
depends on the interaction between the particles or the magnetic clusters.
In the case of magnetic clusters the interactions between the particles
are weak and hence the sensitivity is strong. On the other hand, in
a normal ferromagnetic and antiferromagnetic systems, the interaction
between the magnetic atoms is strong and sufficiently large frequencies
(usually MHz or GHz) are needed to see any significant amount of shift
in the frequency-dependent \textit{ac} susceptibility curves. \cite{Mydosh relative shift in Tf,Spin-glass behavior in Ni-doped La1.85Sr0.15CuO4}

\begin{figure}
\includegraphics[scale=0.35]{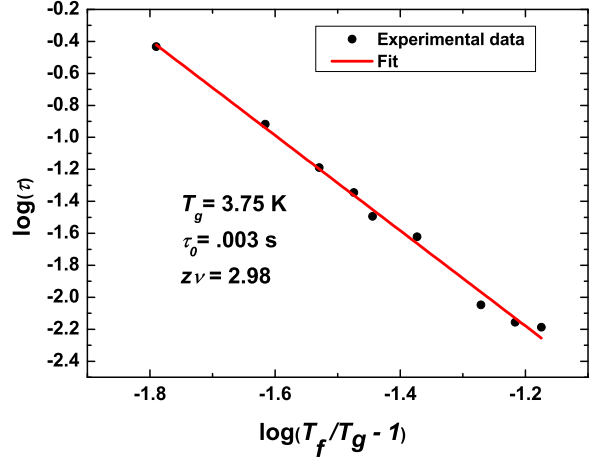}\caption{\label{fig:7Tf fit}The best fit of spin freezing temperatures ($T_{f}$)
to the equation $\tau/\tau_{0}=(T_{f}/T_{g}-1)^{-z\nu}$}
\end{figure}

In Fig. \ref{fig:7Tf fit} we have shown the fit to the conventional
expression of the critical slowing down of the relaxation times, $\tau/\tau_{0}=(T_{f}/T_{g}-1)^{-z\nu}$
. In this expression, the characteristic time $\tau$ represents the
dynamical fluctuation time scale and corresponds to the observation
time ($t{}_{obs}=1/\omega=1/2\pi\nu)$, $\tau_{0}$ is the microscopic
relaxation time, $T_{f}$ is the freezing temperature at a specific
observation time, $T_{g}$ is the spin glass transition temperature
which is equivalent to $T_{f}$ as $\nu\rightarrow0$, and $z\nu$
is the dynamical exponent. The best fit is obtained for $T_{g}\approx3.75$
K, $z\nu\approx2.98$ and $\tau_{0}\approx3\times10^{-3}$ s. For
a conventional spin glass system the $\tau_{0}$ value lies between
$10^{-10}$ and $10^{-13}$ s (Ref.\cite{t0 10-13s conventional SG,JPCM Nd5Ge3})
and $z\nu$ ranges from $4$ to $13$. The $\tau_{0}$ value obtained
for Zn$_{\text{3}}$V$_{\text{3}}$O$_{\text{8}}$ is much higher
than that for conventional spin glasses; in fact it is close to the
values obtained in reentrant spin glass (RSG) systems and some other
SG systems like LaCo$_{0.5}$Ni$_{0.5}$O$_{3}$ (Ref.\cite{LaCo0.5Ni0.5O3 PSG PRB Vishwanathan}),
BiFeO$_{3}$ (Ref.\cite{BiFeO3 SG PRB}), and LaMn$_{0.5}$Fe $_{0.5}$O$_{3}$
(Ref.\cite{LaFe0.5Mn 0.5O3 high T0 unusual SG}) where $\tau_{0}$
is close to $10^{-5}$s . Such a high value for $\tau_{0}$ indicates
that the spin flipping takes place in a rather slow manner in Zn$_{\text{3}}$V$_{\text{3}}$O$_{\text{8}}$.
The value of $z\nu$ is also less than the range for usual spin glasses.

\begin{figure}
\includegraphics[scale=0.35]{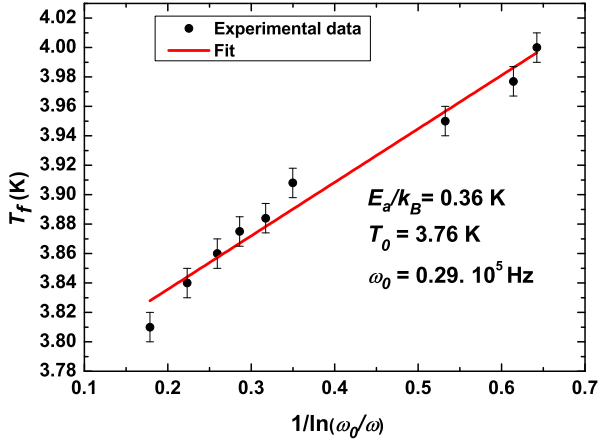}

\caption{\label{fig:8Vogel Fulcher fit}The plot of $T_{f}$ versus $1$/ln($\omega_{0}/\omega$)
fitted with the Vogel-Fulcher law.}
\end{figure}

In fig. \ref{fig:8Vogel Fulcher fit} we have fitted $T_{f}$ with
the empirical Vogel\textendash{}Fulcher law, $\omega=\omega_{0}exp[-E_{a}/k_{B}(T_{f}-T_{0})]$,
where $\omega_{0}$ is the characteristic frequency, $\omega$ is
the angular frequency ($\omega=2\pi\nu$), $E_{a}$ and $T_{0}$ are
the activation energy and Vogel\textendash{}Fulcher temperature, respectively.
The best fit, shown in fig. \ref{fig:8Vogel Fulcher fit}, is obtained
for $\omega_{0}=0.29\times10^{5}$ Hz, $E_{a}/k_{B}=0.36$ K, and
$T_{0}=3.76$ K. The characteristic frequency obtained from the fitting
is much lower than that of conventional spin glass systems, which
is about $10^{13}$ rad/s.\cite{CuMn Mydosg conventional SG,U2IrSi3 high f0 conventional SG}
Such a low value of characteristic frequency is associated with RSG
systems like Ni$_{2}$Mn$_{1.36}$Sn$_{0.64}$. \cite{small f0 sumanta PRB NiMn1.36Sn0.64}
These observations clearly suggest that the spin glass state in Zn$_{\text{3}}$V$_{\text{3}}$O$_{\text{8}}$
is not atomic in origin; rather it is related to clusters of atoms
and hence we would identify it as a cluster glass. 

\subsubsection{\textbf{Relaxation}}

\begin{figure}
\includegraphics[scale=0.35]{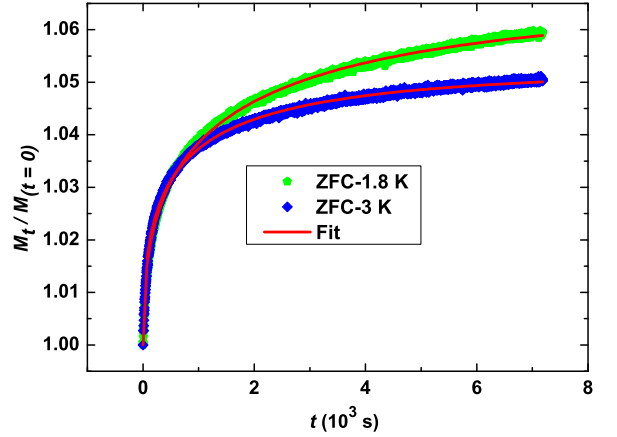}\caption{\label{fig:9ZFC}ZFC magnetization relaxation (normalised with respect
to the magnetization at $t=0$) measured at $T=1.8$ K (green pentagon)
and 3 K (blue diamond) , $H=100$ Oe. The solid lines represent the
fit with the stretched exponential expression given in text.}
\end{figure}

To study the dynamics of the glassy state, we measured the ZFC magnetization
relaxation (see fig. \ref{fig:9ZFC}) using the following protocol:
the sample was cooled in zero field (ZFC) from well above $T_{f}$
to the measuring temperature. Once the measurement temperature was
reached, a field of $10$0 Oe was applied and the magnetization $M_{ZFC}(t)$
was then recorded with time. We can clearly see that the ZFC magnetization
does not saturate even after $2$ h. This is because, in the glassy
state, the moments are randomly frozen and it takes a long time for
the field to turn those spins along the field direction. We have fitted
the time dependence with the standard stretched exponential expression
given below,

\begin{equation}
M_{t}(H)=M_{0}(H)+[M_{\infty}(H)-M_{0}(H)][1-exp\{-(t/\tau)^{\alpha}\}]
\end{equation}

In this formula, $\tau$ is the characteristic relaxation time and
$\alpha$ is the stretching parameter, which ranges between $0$ and
$1$. Here $M_{0}$ and $M_{\infty}$ are the magnetization values
at $t\sim$$0$ and $t\sim\infty$, respectively. The best fit is
obtained for using $M_{\infty}/M_{0}$= $1.07$ and $1.05$, $\tau$
= $1402$ s and $627$ s, $\alpha$ = $0.46$ and $0.45$ for $T=1.8$
K and $3$ K respectively. Note that the growth of the magnetization
is slower at $1.8$ K compared to $3$ K since at $1.8$ K the system
is deeper in the frozen state. However, no significant relaxation
is observed in the FC case (not shown in Fig. \ref{fig:9ZFC}). The
value of $\tau$ obtained for Zn$_{\text{3}}$V$_{\text{3}}$O$_{\text{8}}$
is a bit smaller compared to those obtained in other spin glass systems
like Nd$_{5}$Ge$_{3}$(Ref.\cite{JPCM Nd5Ge3}) and U$_{\mbox{2}}$PdSi$_{3}$(Ref.\cite{U2PdSi3 MIRM PRB}).
The reason is that the value of $\tau$ depends on how deep into the
frozen state the measurement has been carried out. In the above mentioned
systems the $T_{f}$ is much higher ($30$ K for Nd$_{5}$Ge$_{3}$
(Ref.\cite{JPCM Nd5Ge3}) and $13.5$ K for U$_{\mbox{2}}$PdSi$_{3}$
(Ref.\cite{U2PdSi3 MIRM PRB})) compared to that in Zn$_{\text{3}}$V$_{\text{3}}$O$_{\text{8}}$.
So, when the measurements are carried out at $12$ K (for Nd$_{5}$Ge$_{3}$)
or at $2$ K, $5$ K (for U$_{\mbox{2}}$PdSi$_{3}$) the system is
already deep into the frozen state while for Zn$_{\text{3}}$V$_{\text{3}}$O$_{\text{8}}$,
even at $1.8$ K the system is not that deep into the frozen state
compared to those two systems. Indeed, the value of $\tau$ for Zn$_{\text{3}}$V$_{\text{3}}$O$_{\text{8}}$
($T_{f}$ = $3.75$ K) at $1.8$ K is $1402$ s which is similar to
what is obtained for U$_{\mbox{2}}$PdSi$_{3}$($T_{f}$ = $13.5$
K) at $10$ K ($1411$ s).\cite{U2PdSi3 MIRM PRB}

\begin{figure}
\includegraphics[scale=0.35]{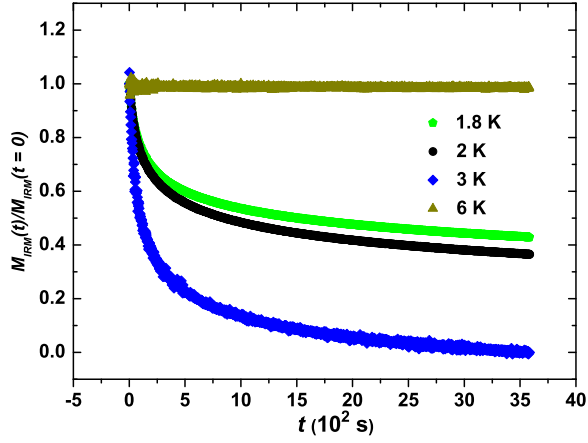}\caption{\label{fig:10IRM}Isothermal remanent magnetisation normalised with
respect to the value at $t=0$ is plotted as a function of time at
$1.8$ K (green pentagon), $2$ K (black circle), $3$ K (blue diamond)
and $6$ K (dark yellow triangle). }
\end{figure}

We have also measured the isothermal remanent magnetization ($M_{IRM}$)
of Zn$_{\text{3}}$V$_{\text{3}}$O$_{\text{8}}$ to explore the metastable
behavior of the glassy state at temperatures below the spin glass
transition temperature (see Fig. \ref{fig:10IRM}). For this, first
we cooled the sample in the ZFC mode from $150$ K to the desired
temperature, then a field of $40$ kOe was applied for $300$ s and
then the applied field was switched off. The magnetization was then
recorded as a function of time for 1 hour. The decay of the remnant
magnetization with time is significantly slow in the spin glass state.
It is natural that the decay of $M_{IRM}$ is slower at lower temperatures
(where the system is deeper into the frozen state) and gets faster
as one gets closer to $T_{f}$. This indicates that the application
of a field below $T_{f}$ causes the system to go to a metastable
and irreversible state.\cite{JPCM Nd5Ge3,U2PdSi3 MIRM PRB,U2IrSi3 high f0 conventional SG}
As expected, above $T_{f}$ $M_{IRM}$ is independent of time. Isothermal
remnant magnetization($M_{IRM}$) data could also be fit using eqn.1
and the fit gives similar parameters as obtained from the ZFC magnetic
relaxation data.

\subsubsection{\textbf{Aging}}

\begin{figure}

\includegraphics[scale=0.35]{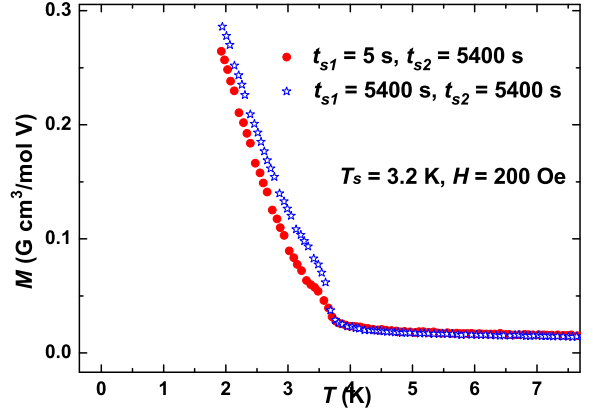}\caption{\label{fig:11 aging 1}Thermoremanent magnetization (TRM) versus temperature
for $T_{s}=3.2$ K , $H=200$ Oe after using different stopping times
during cooling.}

\end{figure}

The nonergodic metastable state of Zn$_{\text{3}}$V$_{\text{3}}$O$_{\text{8}}$
is further verified by experiments manifesting the aging phenomena.
To observe the aging effects, thermoremanent magnetization (TRM) was
measured using the following protocol; the sample was first cooled
in a field of $200$ Oe from a temperature well above $T_{f}$ to
a stop temperature, $T_{s}=\mbox{3.2}$ K (below $T_{f}$) where the
system was allowed for a stop time $t_{s1}$ without changing the
field. Then, the magnetic field was switched off and a second stop
time $t_{s2}$ was provided. The magnetic field was then set back
to the original value ($200$ Oe) and the sample was immediately cooled
down to the lowest temperature ($1.8$ K). Subsequently, the field
was switched off and the magnetization was measured as a function
of temperature during the heating. The measurement was performed for
two different values of $t_{s1}$, namely $5$ and $5400$ s. The
second stop time $t_{s2}$ was kept constant ($5400$ s) for both
the measurements. The salient feature of this measurement is that
the two curves bifurcate at a temperature ($3.66$ K) slightly below
$T_{f}$ (see fig. \ref{fig:11 aging 1}), indicating the influence
of the aging on the spin glass state. Similar behavior has been also
seen in La$_{0.95}$Sr$_{0.05}$CoO$_{3}$ (Ref.\cite{t0 10-13s conventional SG}),
Nd$_{5}$Ge$_{3}$ (Ref.\cite{JPCM Nd5Ge3}), and Ag(Mn) (Ref.\cite{AgMn Aging effect})
spin glass systems. 

\begin{figure}
\includegraphics[scale=0.35]{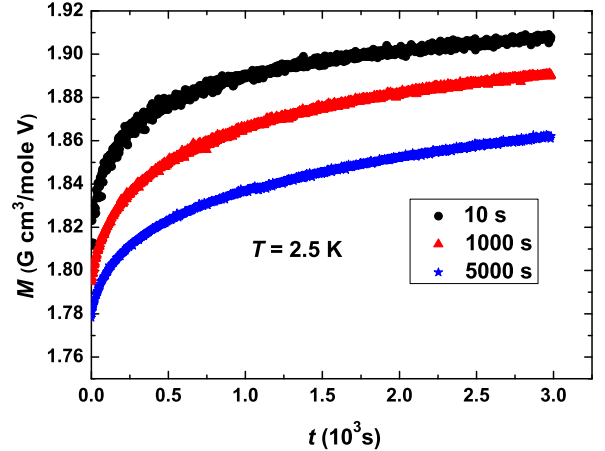}\caption{\label{fig:12 aging 2} Aging effect is manifested by the growth of
the magnetization as a function of time. The sample was cooled (ZFC)
to $2.5$ K and the field ($200$ Oe) was applied after different
time intervals of $5$, $1000$ and $5000$ s.}
\end{figure}

Fig.\ref{fig:12 aging 2} shows the growth of the magnetization data
as a function of time, in the metastable state. The sample was cooled
to $2.5$ K in the ZFC mode and a field of $200$ Oe was applied after
a waiting time $t$. The magnetization was then measured as a function
of time. As we can see, the magnetization growth is slower for larger
waiting time, which again points that the metastability associated
with the low temperature magnetic state.

\subsubsection{\textbf{Memory effect}}

\begin{figure}

\includegraphics[scale=0.35]{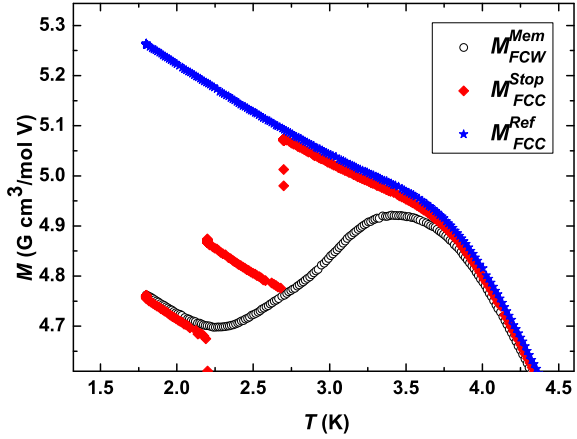}\caption{\label{fig:13Memory effect Zn3V3O8}The memory effect in Zn$_{\text{3}}$V$_{\text{3}}$O$_{\text{8}}$
is observed in the temperature variation of FC magnetization. Here
$M\protect\overset{Stop}{_{FCC}}$ curve was obtained during cooling
the sample in a field of $500$ Oe with intermediate stops of $3$
h duration each at $2.7$ K and $2.2$ K. At each stopping temperature,
the field was switched off. $M\protect\overset{Mem}{_{FCW}}$ was
measured during continuous heating of the sample in the same field.
The reference curve ($M\protect\overset{Ref}{_{FCC}}$) was measured
during continous cooling of the sample in the same field ($H=500$
Oe).}

\end{figure}

To observe the existence of a memory effect\cite{memory effect in nanoparticle}
in the \textit{dc} magnetization we have measured the FC magnetization
using the following protocol. The magnetization was recorded during
cooling of the sample at $500$ Oe from $100$ K down to $2$ K at
a constant cooling rate of $1$ K/min. The cooling process was interrupted
at $2.7$ K and $2.2$ K for a waiting time $t_{w}$ = $3$ h, in
each case. During $t_{w}$, the field was switched off and the system
was allowed to relax. After each stop and wait period, the FC process
was resumed. The stops at $2.7$ K, and $2.2$ K are evident in the
obtained $M\overset{Stop}{_{FCC}}$ curve in Fig. \ref{fig:13Memory effect Zn3V3O8},
as step like features. Once the cooling process was completed by reaching
$2$ K, the sample was heated continuously in the same magnetic field
and heating rate while recording the magnetization data. The magnetization
obtained this way, referred to as $M\overset{Mem}{_{FCW}}$, exhibits
a weak change of slope at $2.7$ K and a prominent minimum at $2.2$
K although there is no stop at these temperatures. This indicates
that the system has its previous behavior during the cooling operation
imprinted as a memory. This sort of behavior has been observed in
intermetallic compounds such as GdCu (Ref.\cite{memory effect in GdCu}),
Nd$_{5}$Ge$_{3}$(Ref.\cite{JPCM Nd5Ge3}) and in superspin glass
nanoparticle systems\cite{memory effect in nanoparticle,memory effect in Fe3N nanoparticle }
This is considered to be a typical characteristic of spin glasses.
The dip at $2.7$ K in the $M\overset{Mem}{_{FCW}}$ curve is weak
because at $2.7$ K the system is not much below the blocking temperature
($T_{b}=3.8$ K at $H=500$ Oe) which is the peak of the ZFC curve
(see Fig. \ref{fig:3ZFC-FC}). This signifies that at $2.7$ K the
system is not deep enough into the SG state. A reference curve ($M\overset{Ref}{_{FCW}}$)
was also measured by simply cooling the sample continuously in $H=500$
Oe. No memory effect has been observed when we waited at $15$ K,
a temperature above $T_{b}$ (not shown in Fig. \ref{fig:13Memory effect Zn3V3O8}).

\begin{figure}

\includegraphics[scale=0.33]{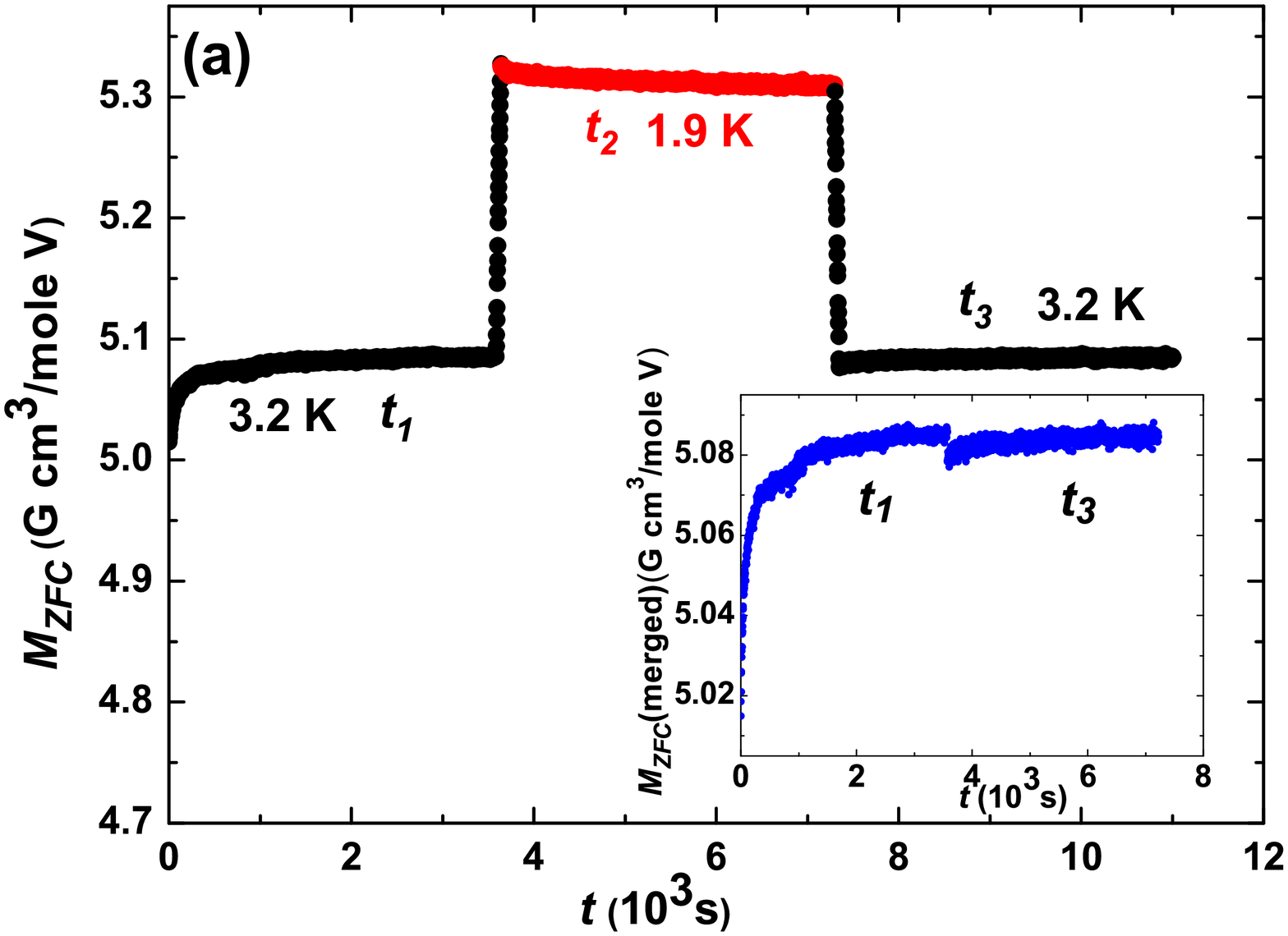}

\includegraphics[scale=0.33]{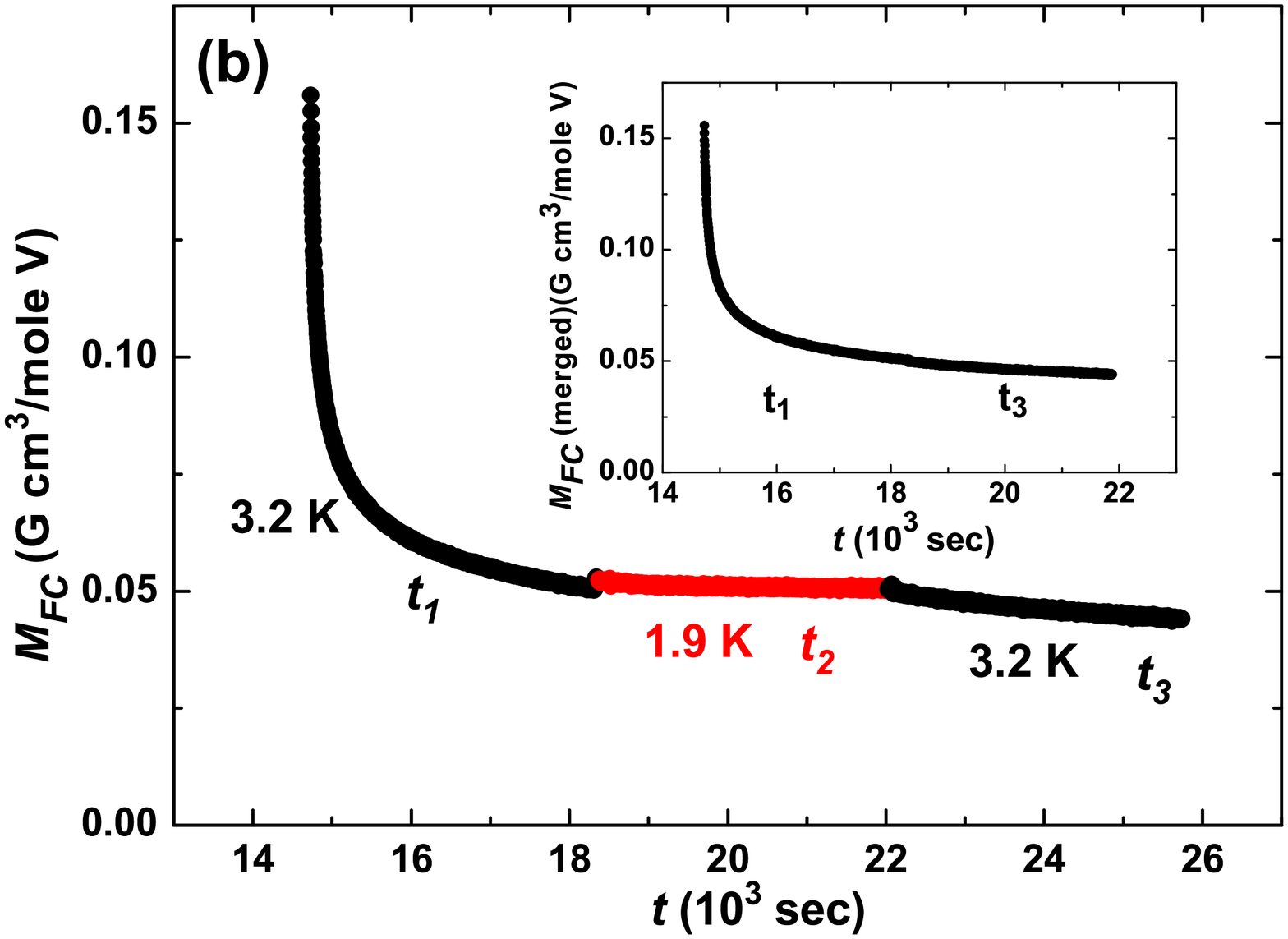}\caption{\label{fig:14 Memory ZFC FC}Magnetic relaxation in Zn$_{\text{3}}$V$_{\text{3}}$O$_{\text{8}}$
at $3.2$ K with a temporary quench to $1.9$ K for (a) the ZFC method
and (b) the FC method. Insets show the relaxation data during $t_{1}$
and $t_{3}$ merge nearly perfectly.}
\end{figure}

To further test the signature of the memory effect we have investigated
the ZFC and FC relaxation behavior with negative $T$ cycling as shown
in Fig. \ref{fig:14 Memory ZFC FC}. In the ZFC method, the sample
was first zero field cooled down from the paramagnetic phase to the
measuring temperature $T_{1}=3.2$ K, which is below the spin freezing
temperature $T_{f}$ . Subsequently, a magnetic field of $500$ Oe
was applied and the magnetization was recorded as a function of time
for a time period $t_{1}=1$ h. After that, the sample was quenched
to a lower temperature $T_{2}=1.9$ K without changing the field and
the magnetization was recorded for a time $t_{2}=1$ h . Finally,
the temperature was restored to $T_{1}=3.2$ K and the magnetization
was recorded for a time $t_{3}=1$ h. The relaxation curve obtained
this way is depicted in Fig. \ref{fig:14 Memory ZFC FC}(a). When
the system was returned to $T_{1}=3.2$ K after the temporary quenching,
the magnetization resumes from the previous value it reached before
the temporary quenching. This indicates that the temporary quenching
does not erase the memory in ZFC relaxation. In the FC process, the
sample was first field cooled to $T_{1}=3.2$ K in $500$ Oe. Once
the measuring temperature was reached, the field was switched off
and subsequently the magnetization was measured as a function of time
(see Fig. \ref{fig:14 Memory ZFC FC}(b)). Similar to the ZFC method,
the FC method also preserves the state of the system even after a
temperature quench. In both ZFC and FC methods, the relaxation curve
during $t_{3}$ is just a continuation of the curve during $t_{1}$
as shown in the insets of Fig. \ref{fig:14 Memory ZFC FC}, which
represents a memory effect.

\begin{figure}

\includegraphics[scale=0.33]{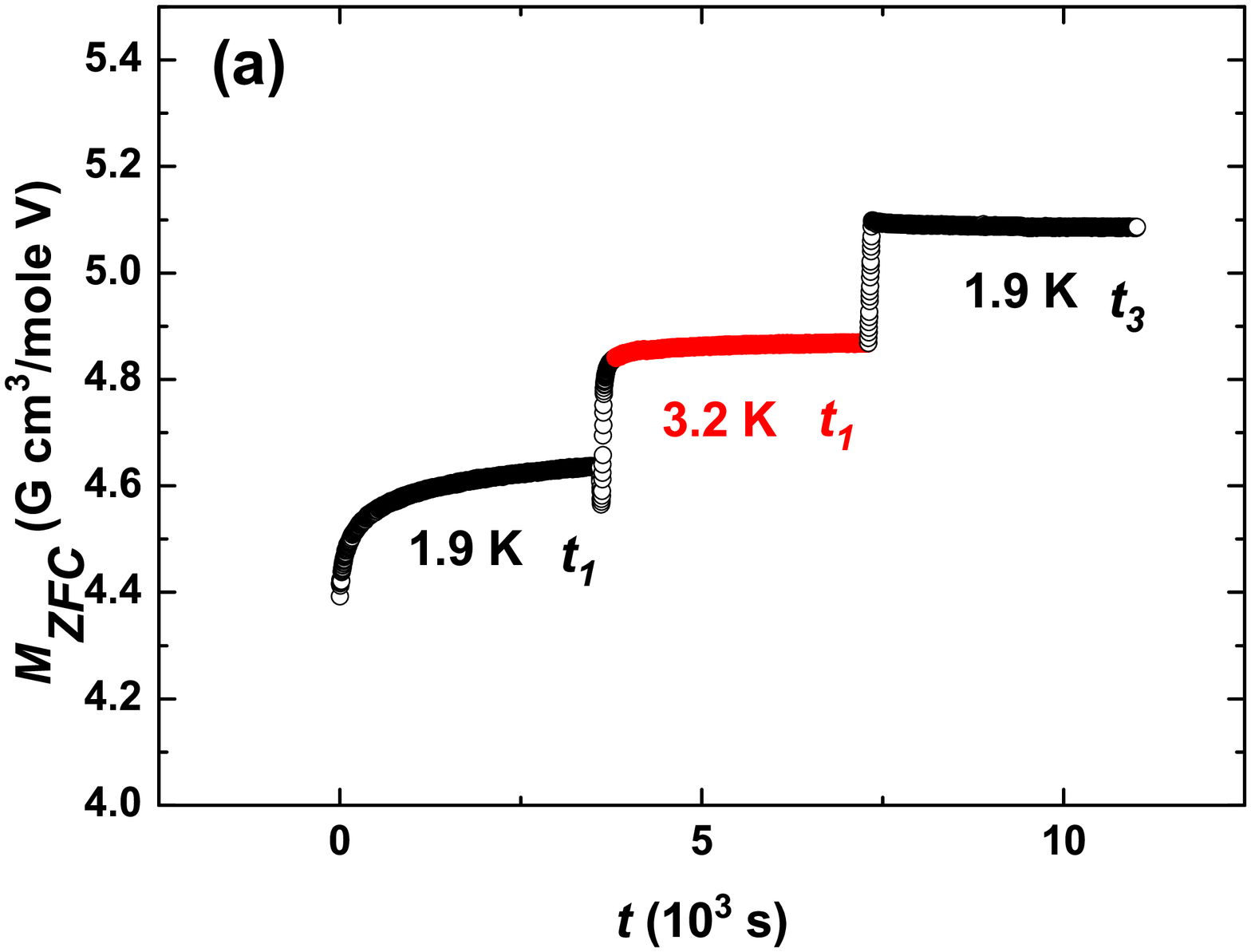}

\includegraphics[scale=0.33]{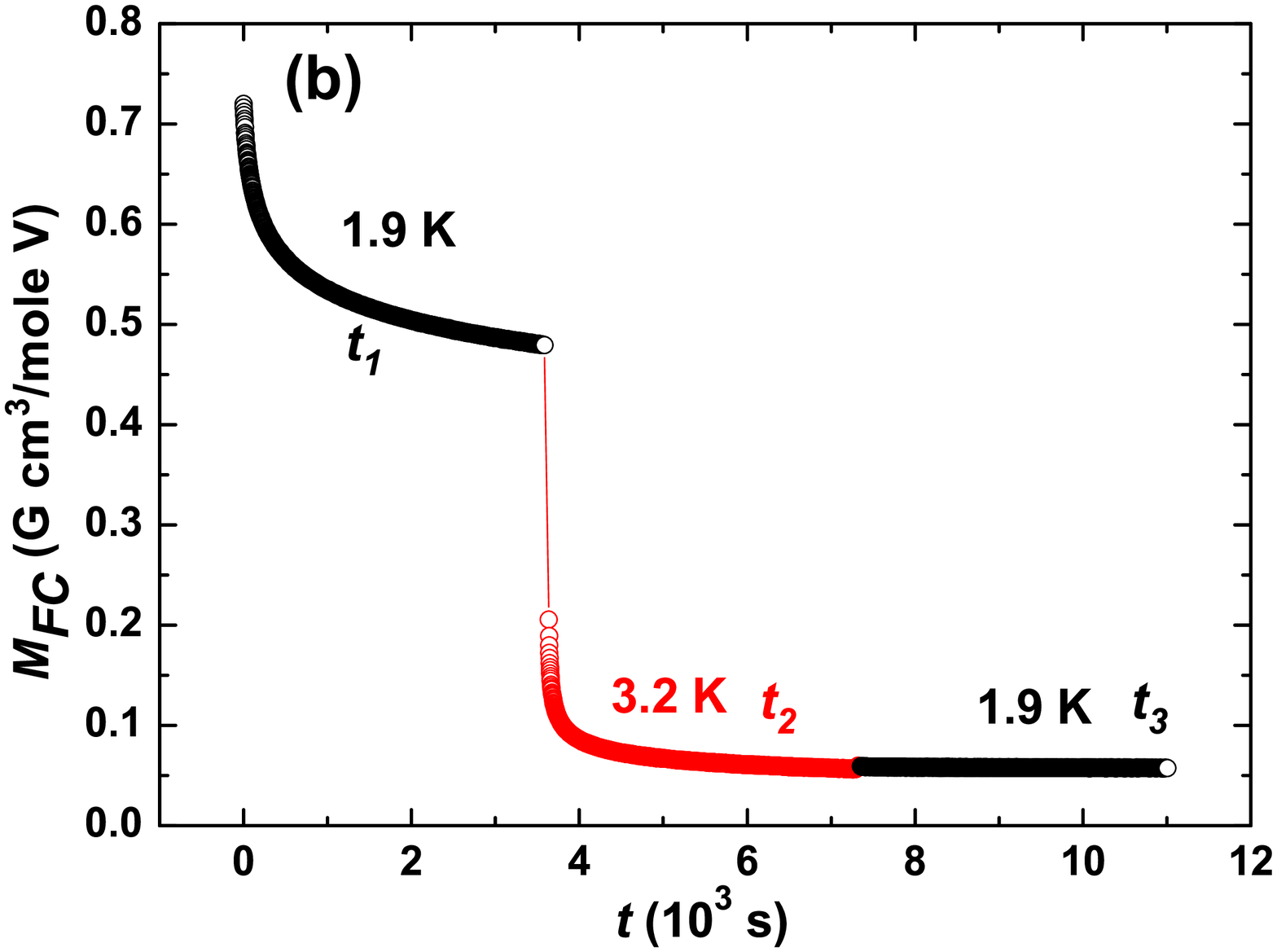}

\caption{\label{fig:15 positive heat cycle}Magnetic relaxation in Zn$_{\text{3}}$V$_{\text{3}}$O$_{\text{8}}$
with a positive heating cycle for (a) ZFC method and (b) FC method.}
\end{figure}

According to the droplet model \cite{droplet model 1 Fischer,droplet model 2 Fischer}
of spin glasses, one would expect a symmetric behavior in magnetic
relaxation with respect to heating and cooling. But the hierarchical
model \cite{Hierarchical model EPL,memory effect in nanoparticle}
predicts that a positive temperature cycle can destroy the previous
memory and initialize the relaxation again, which means the response
is asymmetric. In order to compare the response with respect to intermittent
heating and cooling cycles, the relaxation experiment with a temporary
heating cycle was also performed. The results are shown in Fig. \ref{fig:15 positive heat cycle}(a)
and (b). As can be seen from these figures, a positive temperature
cycling erases the memory and reinitializes the relaxation in both
ZFC and FC processes. This clearly suggests that the response of the
system is asymmetric, therefore it supports the hierarchical picture
proposed for spin glasses. According to this model, there exists a
multi-valley free-energy surface of a frustrated system at a given
temperature $T$. When the temperature of the system is lowered from
$T$ to $T-\Delta T$, then each valley is split into many sub-valleys.
If $\Delta T$ is large the energy gaps between the primary valleys
are also large, and the system cannot overcome this energy barrier
within a given time $t_{2}$. So, the relaxation occurs only within
the secondary sub-valleys. When the system is brought back to its
intial temperature temperature $T$, the sub-valleys coalesce back
to the original free-energy surface and relaxation at $T$ resumes
without being perturbed by the intermediate relaxations at $T-\Delta T$.
But when the system is heated from $T$ to $T+\Delta T$ , then the
barriers between the free energy primary valleys are lowered or sometimes
even merge. Therefore, the relaxations can easily take place within
different primary valleys. When the temperature $T$ is restored,
although the free energy surface goes back to the original free energy
surface, the relative occupancy of each primary energy valley does
not remain the same as before. Therefore, the state of the system
changes after a temporary heating cycle and no memory effect is observed.
Behavior of the kind observed in Zn$_{\text{3}}$V$_{\text{3}}$O$_{\text{8}}$
has been seen in some other SG systems too.\cite{JPCM Nd5Ge3,memory effect in GdCu,memory effect in nanoparticle}

\subsection{Heat capacity}

\begin{figure}
\includegraphics[scale=0.3]{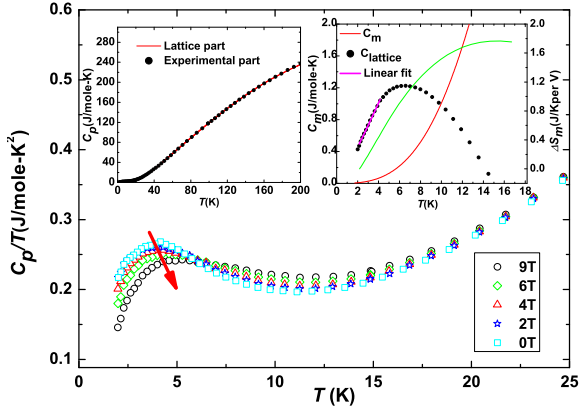}

\caption{\label{fig:16heat capacity}(Left inset) The temperature dependence
of specific heat of Zn$_{\text{3}}$V$_{\text{3}}$O$_{\text{8}}$;
the red line represents the fitting of the heat capacity using one
Debye and two Einstein terms (see text). The right inset displays
the magnetic contribution of specific heat at zero field, the red
line indicates the lattice contribution of heat capacity, the magenta
line shows the fit with the power law (given in text); the green line
(right axis, left inset) shows the change of entropy calculated by
integrating the $C_{\text{m}}/T$ data. In the main figure $C_{P}/T$
vs. $T$ is shown at zero field (cyan square), $20$ kOe (blue star),
$40$ kOe (red triangle), $60$ kOe (green diamond) and $90$ kOe
(black circle) to depict the slight shift of the maximum of $C_{P}/T$
curve with the increase in magnetic field. Note that in both the insets
and in the main figure the heat capacity is calculated for one formula
unit. }
\end{figure}

The heat capacity data of Zn$_{\text{3}}$V$_{\text{3}}$O$_{\text{8}}$
are shown in Fig. \ref{fig:16heat capacity}. It is clear from the
figure that no sign of long-range ordering is observed down to 2 K.
A small hump is seen in the $C_{\text{p}}/T$ data, at around $3.75$
K, which shifts slightly with the increase in the magnetic field.
However this change with magnetic field is not due to the Schottky
anomaly. 

Since we do not have a suitable non-magnetic analogue for this system
we tried to extract the magnetic specific heat of Zn$_{\text{3}}$V$_{\text{3}}$O$_{\text{8}}$
by subtracting the lattice contribution using a combination of Debye
and Einstein heat capacities, $C_{Debye}$ and $C_{Einstein}$, respectively.
In the $T$-range $30-110$ K, the measured heat capacity $C_{P}$
could be fit with a combination of one Debye and two Einstein functions
of the type given below where the coefficient $C_{d}$ stands for
the relative weight of the acoustic modes of vibration and the coefficients
$C_{e_{1}}$ and $C_{e_{2}}$ are the relative weights of the optical
modes of vibration. 

$C_{Debye}=\ensuremath{C_{d}\times9nR}\ensuremath{(T/\theta_{\text{d}})^{\mbox{\text{3}}}}\ensuremath{\int_{\text{0}}^{\text{\ensuremath{\theta\ensuremath{_{\text{d}}}/T}}}(x^{\text{4}}e^{\text{\ensuremath{x}}}/(e^{\text{\ensuremath{x}}}-1)^{\text{2}}}\ensuremath{)dx}$

$C_{Einstein}=\ensuremath{3nR[\sum C{}_{e_{m}}\times\frac{x_{E_{m}}^{2}e^{x_{E_{m}}}}{(e^{x_{E_{m}}}-1)^{2}}]}$,
$x=\frac{h\omega_{E}}{k_{B}T}$

In the above formula, $n$ is the number of atoms in the primitive
cell, $k_{\text{B}}$ is the Boltzmann constant, and $\theta_{\text{d}}$
is the relevant Debye temperature, $m$ is an index for an optical
mode of vibration. In the Debye-Einstein model the total number of
modes of vibration (acoustic plus optical) is equal to the total number
of atoms in the primitive unit cell. For Zn$_{\text{3}}$V$_{\text{3}}$O$_{\text{8}}$
this number is $14$. In this model we have considered the ratio of
the relative weights of acoustic modes and sum of the different optical
modes to be $1:n-1$. Due to having two heavy atoms (vanadium and
zinc) and one comparatively lighter atom (oxygen) in this compound
we considered two different optical modes of vibrations. The fit yields
a Debye temperature of 127 K and Einstein tempertures of $249$ K
and $600$ K with relative weights $C_{d}:C_{e_{1}}:C_{e_{2}}$ $=$$2.6:12:19$.
Upon subtracting the lattice heat capacity with the above parameters,
we obtain the magnetic contribution to the heat capacity $C_{\mathrm{m}}(T)$.
The entropy change ($\Delta S)$ was calculated by integrating the
$C_{m}/T$ data (see Fig. \ref{fig:16heat capacity}). The entropy
change from about 14 K to 2 K is about 1.76 J/K (calculated for one
formula unit containing three vanadium ions). For $S=1/2$ and $S=1$
systems the values of the entropy change ($\Delta S=Rln(2S+1)$) are
$5.763$ J/K and $9.314$ J/K respectively. If two vanadiums in Zn$_{\text{3}}$V$_{\text{3}}$O$_{\text{8}}$
are in the 3+ oxidation state ($S=1$) and the remaining one is in
the 4+ state ($S=1/2$) then the total entropy change expected in
case of long range order would be $24.391$ J/K. The value of $\Delta S$
obtained for Zn$_{\text{3}}$V$_{\text{3}}$O$_{\text{8}}$ is only
about 7\% of this value which indicates the presence of many degenerate
low-energy states at low temperatures.\cite{note} This large reduction
in the value of $\Delta S$ down to temperatures much lower than the
Weiss temperature ($\theta_{CW}$) is typical of disordered systems
and a consequence of the presence of strong geometric frustration
in Zn$_{\text{3}}$V$_{\text{3}}$O$_{\text{8}}$. We observed a broad
maximum at 6 K in the $C_{\mathrm{m}}(T)$ vs. $T$ data above the
freezing temperature ($T_{f}$) which is a characteristic feature
of spin-glass systems.\cite{URh2Ge2 Chi" not zero below Tf,Cp peak > Tg}
At low $T$, (in the range of $2-4$ K) $C_{\mathrm{m}}(T)$ varies
nearly as $C_{\mathrm{m}}(T)$ = $\gamma T^{\alpha}$ with $\gamma=0.21$
JK$^{-2.2}$mol$^{-1}$ and $\alpha$= $1.15$, i.e., nearly linear
with temperature. Particularly in SG systems this kind of linear variation
of the low-temperature magnetic specific heat is claimed to be a common
feature\cite{Power law in Cp in SG PRB,Power law in Cp in SG PRB2,Power law in CP in SG PRL}.
All these facts collectively point towards the formation of a metastable
frozen state in the system at low temperature.

\section{Conclusion}

In this work we have reported the crystal structure, \textit{dc} and
\textit{ac} magnetization, various static and dynamic magnetic measurements,
and heat capacity of a new vanadium-based multivalenced spinel (AB$_{2}$O$_{4}$)
system Zn$_{\text{3}}$V$_{\text{3}}$O$_{\text{8}}$. The $\chi(T)$
data evidence the presence of strong antiferromagnetic correlations
between the magnetic ions. The Curie constant is consistent with the
presence of two vanadiums in the $3+$ oxidation state and one in
the $4+$ state in each formula unit. Below about $3.75$ K (well
below $\theta_{CW}$ of $-370$ K), ZFC and FC magnetization curves
bifurcate from each other which is suggestive of spin glass behaviour.
This is supported by the measurements of \textit{ac} susceptibility
where we observed a peak at about $3.75$ K (both in $\chi'$ and
$\chi''$) and the freezing temperature has a logarithmic variation
with the change in the measuring frequency. Below $T_{g}$, $\chi''$attains
a constant value which again provides a signature of SG behavior in
Zn$_{\text{3}}$V$_{\text{3}}$O$_{\text{8}}$. From the value of
the characteristic frequency ($\omega_{0}=0.29\times10^{5}$ rad/s)
obtained from the Vogel-Fulcher fit, we conclude that the system is
closer to a cluster spin glass where the magnetic entities involved
are bigger than at the atomic level. We also observed the relaxation
of isothermal remanent magnetization and the growth of the ZFC magnetization,
which are very commonly observed in SG systems. Further, ageing phenomena
and memory effect were also observed both in ZFC and FC magnetization.
We observed that a positive temperature cycle erases the memory while
a negative temperature cycle retains memory. This type of behavior
is predicted by the hierarchical model of SG systems. In the heat
capacity measurement we did not observe any sharp anomaly indicative
of long range ordering down to $2$ K. At around $3.75$ K a small
hump is seen in the $C_{\text{p}}/T$ data. The entropy change $\lyxmathsym{\textgreek{D}}S$
is only about $7$ \% of what is expected for an ordered system containing
two $S=1$ and one $S=1/2$ ions per formula unit. This is likely
due to the presence of strong geometric frustration in the system.
The magnetic heat capacity shows an almost linear variation below
$T_{f}$, which is another common feature in SG systems. All these
above features point to the formation of a metastable, nonergodic
state in Zn$_{3}$V$_{3}$O$_{8}$ below $3.75$ K.

\section{Acknowledgement}

Discussions with B. Koteswararao are acknowledged. The authors thank
the Department of Science and Technology, Govt. of India for financial
support.


\begin{thebibliography}{References}
\bibitem[39]{note}Depending on the temperature range chosen for the
fit, the inferred lattice heat capacity is a bit different. Also,
since this system has a large $\theta_{CW}$ (= -370 K), entropy release
might take place in a broad temperature range. Due to the large lattice
contribution at high temperatures it is nearly impossible to accurately
determine the magnetic contribution to the heat capacity in this range.
Consequently, there is likely to be a significant uncertainty in the
entropy change estimation.
\end{thebibliography}
\end{document}